\documentclass[ reprint, amsmath,amssymb,aps]{revtex4-1}

\pdfoutput=1
\usepackage{epsfig,graphicx,xcolor,amsbsy,amssymb,latexsym,amsfonts,amsmath,setspace}
\usepackage{pstricks}
\usepackage{color}

\usepackage{eurosym}
\usepackage{graphicx}
\usepackage{tikz}
\usepackage{xcolor}
\usepackage{amsmath,amssymb,amsfonts,pstricks,setspace}
\newcommand{\bea}{\begin{eqnarray}\displaystyle}
\newcommand{\eea}{\end{eqnarray}}

\newcommand{\sq}[2]{\mbox{{\raisebox{.06in}{$#1$}}}\underset{\mbox{$#2$}}{\mbox{\LARGE{$\square$}}}}
\vspace{-0.5cm}
\begin{document}

\title{
%\begin{flushright}{\vspace{-2.5cm}\small SNUST 15-07\\}\end{flushright}
\vspace{-1.8cm}
\bf{Bound States of Little Strings and Symmetric Orbifold CFTs}\\[15pt]}
\author{Ambreen Ahmed$^{\dagger}$,~Stefan~Hohenegger$^{\S\ddagger}$,~Amer Iqbal$^{\dagger\ddagger\star}$,~ Soo-Jong Rey$^{\ddagger\diamond}$
\\
}
\date{\today}
\affiliation{${}^{\dagger}$ Abdus Salam School of Mathematical Sciences, G.C. University, Lahore {\rm PAKISTAN} \\
${}^{\S}$ Universit\'e de Lyon UMR 5822, CNRS/IN2P3, Institut de Physique Nucl\'eaire de Lyon, 4 rue Enrico Fermi, 69622 Villeurbanne Cedex {\rm FRANCE}\\
${}^{\ddagger}$ Fields, Gravity \& Strings, CTPU, Institute for Basic Sciences, Daejeon 34047 {\rm KOREA}
\\
${}^{\star}$ Center for Theoretical Physics, Lahore, {\rm PAKISTAN}
\\
${}^{\diamond}$ School of Physics and Astronomy \& Center for Theoretical Physics, Seoul National University, Seoul 08826  {\rm KOREA}}

\begin{abstract}
We study BPS bound states of little strings in a limit where they realise monopole strings in five dimensional gauge theories. The latter have gauge group $U(M)^N$ and arise from compactification of $(1,0)$ little string theories of type $A_{M-1} \times A_{N-1}$. We find evidence that the partition function of a certain subclass of monopole strings of charge $(k,\ldots,k)$ ($k\geq 1$) is expressible as the partition function of a symmetric orbifold sigma model, whose target space is precisely the symmetric product of the moduli space of monopoles with charge $(1, \ldots, 1)$.
\end{abstract}

\maketitle

%%%%%%%%%%%%%%%%%%%%%%%%%%%%%%%%%%%%%%
Little string theories (LSTs) refer to a quantum theory of noncritical strings without gravity, living in six space-time dimensions. It has been noted that their properties are similar to those of strings encountered in other physical applications, such as hadronic strings in QCD and magnetic flux tubes in superconductors. It has also been noted that little string theories provide microstate descriptions of various supersymmetric black holes in four dimensions. When compactified  to five dimensions, the theory gives rise to monopole strings \cite{Bak:2014xwa, Haghighat:2015coa,Hohenegger:2015cba}, whose various properties can be studied using the underlying little strings \cite{Hohenegger:2015btj}. Monopole strings, as the name indicates, are string like solutions in five dimensions which appear pointlike and carry monopole charge in the transverse three dimensions \cite{Boyarsky:2002ck,Douglas:2010iu}. All these connections motivate a deeper study of the structure of LSTs.

LST of type $A_{N-1}\times A_{M-1}$ with ${\cal N} = (1,0)$ supersymmetry can be engineered using $N$ M5-branes probing a transverse $A_{M-1}$ orbifold geometry. The M-theory background is given by $\mathbb{R}^{4}\times \mathbb{T}^{2}\times \mathbb{S}^{1}\times \mathbb{R}^{4}/\mathbb{Z}_{M}$, where the M5-branes are extended along $\mathbb{R}^4\times \mathbb{T}^2$ and separated along $\mathbb{S}^1$. In a series of papers \cite{Hohenegger:2016eqy, Haghighat:2013gba, Hohenegger:2013ala,Hohenegger:2015btj}, such LSTs of type $A_{N-1}$ probing a $\mathbb{Z}_{M}$ orbifold background were studied and their partition functions were calculated by using a dual setup of D5- and NS5-branes in type IIB string theory. The latter in turn is dual to a particular class of toric Calabi-Yau threefolds $X_{N,M}\sim X_{1,1}/\mathbb{Z}_M\times \mathbb{Z}_N$ where $X_{1,1}$ resembles the resolved conifold near certain boundaries of the moduli space. The partition function of the little strings is refined by the insertion of three $U(1)$ currents: $U(1)_{m}$ corresponding to a rotation of the transverse $\mathbb{R}^4$ \cite{Haghighat:2013gba} and $U(1)_{\epsilon_{1,2}}$ acting on $\mathbb{R}^4 \sim \mathbb{C}^2$ as $(z_1,z_2)\mapsto (e^{i\epsilon_1}z_1,e^{i\epsilon_2}z_2)$. When one of the worldvolume directions is compactified, the corresponding five-dimensional worldvolume theory on the M5-branes becomes a gauge theory with gauge group $U(M)^N$ broken down to $U(1)^{NM}$. The five-dimensional theory on $\mathbb{R}^3\times \mathbb{T}^2$ contains monopole strings coming from M2-branes stretched between the M5-branes \cite{Haghighat:2015coa,Hohenegger:2015cba} and wrapped on $\mathbb{T}^2$. Thus, the partition function of little strings calculates the ${\cal N} = (2, 0)$ elliptic genus of monopole strings, in the limit $\epsilon_2\mapsto 0$ which is required by the compactification of the corresponding direction on which $U(1)_{\epsilon_2}$ acts. These monopole strings carry charges $k_{i}~|~i=1,\ldots, N$ (which is the number of M2-branes between the $i$-th and $i+1$-th M5-brane) as well as fractional momenta $p_{a}~|~a=1,\cdots,M$ along the transverse $\mathbb{S}^1$.

In this paper, we study a subsector of the BPS spectrum consisting of states such that $k_{1}=\ldots=k_{N}=k$ and $p_{1}=\ldots=p_{M}=p$. Namely, this subsector %has the special property that %on the string worldsheet the supersymmetry enhances from ${\cal N}=(2,0)$ to ${\cal N} =(2,2)$. This is attributed to the fact that $k$ M2-branes start and end on the same M5-brane, which simply rephrases that $k$ M2-branes are stretched between every adjacent M5-branes. This configuration
with charges $(k,\ldots,k)$ contains $k$ M2-branes starting and ending on any given M5-brane.
%, wrapping around $\mathbb{S}^1$, and then ending on the same M5-brane.
We show that the degeneracies of these BPS states with quantum numbers $(k,p)$ only depend on the product $k\,p$ and therefore degeneracies of higher charge monopole strings of this type are completely determined by degeneracies of those with charge one. We further argue that the partition function of these monopole strings can be expressed in terms of a conformal field theory (CFT) whose target space is a symmetric orbifold.
% conformal field theory (CFT).
%It should be noted, however, that this simple description of a subsector of the BPS spectrum only holds for monopole strings and not for the full little strings.

The partition function of this class of LSTs, which we denote by $Z_{N,M}$, is given by \cite{Haghighat:2013tka,Hohenegger:2013ala,Hohenegger:2015cba}
\begin{align}
Z_{N,M}=(W({\bf T},m,\epsilon_{\pm}))^{N}\,{\cal Z}_{N,M}({\bf t},{\bf T},m,\epsilon_{\pm})\,,\label{DefPartFct}
\end{align}
where ${\bf t}=(t_{1},\cdots,t_{N})$ with $t_{i}$ (with $i=1,\ldots,N$)  being the separation between the $i$-th and the $(i+1)$-th M5-brane and ${\bf T}=(T_{1},\cdots, T_{M})$ determine the charge of the states with respect to the $\mathbb{Z}_{M}$ orbifold. Furthermore, $m$ and $\epsilon_\pm=\tfrac{\epsilon_1\pm\epsilon_2}{2}$ correspond to the $U(1)'s$  we discussed above, which are required to render $Z_{N,M}$ well-defined.

The partition function (\ref{DefPartFct}) receives contributions from two parts. The first part $W({\bf T},m,\epsilon_{\pm})$ is the contribution of a single M5-brane wrapped on a circle with transverse $\mathbb{Z}_{M}$ orbifold. This contribution can be determined by studying the modes of a single six-dimensional abelian tensor multiplet on a circle colored by the transverse orbifold. The second factor ${\cal Z}_{N,M}$ is the contribution coming from M2-branes suspended between the M5-branes. Since these M2-branes are the little strings in six dimensions, this second factor contains the little string and monopole string degeneracies and will be the object of study in this paper. It can be expanded in the following fashion
\bea\nonumber
{\cal Z}_{N,M}({\bf t},{\bf T},m,\epsilon_\pm)=\sum_{k_{1}\ldots k_{N}}Q_{1}^{k_{1}}\ldots Q_{N}^{k_{N}}{\cal Z}_{N,M}^{k_{1}\cdots k_{N}}\,,
\eea
where $Q_i=e^{-t_{i}}$ and ${\cal Z}_{N,M}^{k_{1}\cdots k_{N}}({\bf T},m,\epsilon_{1,2})$ captures the BPS states of $k_{i}$ M2-branes suspended between the $i$-th and the $(i+1)$-th M5-brane respectively.

The free BPS energy associated with the partition function (\ref{DefPartFct}) is given by $F_{N,M}=\ln({\cal Z}_{N,M})$. In this letter, however, we study ($\overline{Q}_a = e^{ - T_a}$)
\begin{align}
{\cal F}_{N,M}= \hskip-0.2cm \sum_{k\geq 1,n\geq 0}Q_{\rho}^{k}\,Q_{\tau}^{n} \prod_{i=1}^N \prod_{a=1}^M \oint\frac{dQ_{i}d\overline{Q}_{a}}{Q_{i}^{k+1}\overline{Q}_{a}^{n+1}}\,\mbox{ln}\,{\cal Z}_{N,M}\, , \label{CalFDef}
\end{align}
which only counts contributions from the aforementioned subspace of the BPS Hilbert space in which every M5-brane has the same number of M2-branes starting and ending on it. This reduced free energy ${\cal F}_{N,M}$ can be written as the infinite sum
\bea\nonumber
{\cal F}_{N,M}(\rho,\tau,m,\epsilon_{1,2})=\sum_{n=1}^{\infty}{1 \over n} G_{N,M}(n\rho,n\tau,n\,m,n\epsilon_\pm)\,,
\eea
and only depends on $(\tau,\rho,m,\epsilon_\pm)$, which are sums of all $\mathbf{T}$ and $\mathbf{t}$ respectively:
\begin{align}
&\tau=\frac{i}{2\pi}(T_1+\ldots+T_M)\,&&\text{and} &&\rho=\frac{i}{2\pi}(t_1+\ldots+t_N)\,.\nonumber
\end{align}
Furthermore, $G_{N,M}(\rho,\tau,m,\epsilon_\pm)=\sum_{k\geq 1}Q_{\rho}^{k}\,G_M^{\footnotesize(k,\cdots,k)}$ (where the superscript contains $N$ entries of $k$) captures the BPS degeneracies of $k$ M2-branes stretched between any adjacent pair of the $N$ M5-branes and winding $k$ times around the transverse $\mathbb{S}^1$. These M2-branes carry arbitrary momenta along the $\mathbb{S}^1$ in M5-brane world-volume.

For $N=M=1$, the free energy ${\cal F}_{1,1}$ captures all the BPS states in the theory
\begin{align}
{\cal F}_{1,1}(\tau,\rho,m,\epsilon_\pm)=\sum_{k>0}Q_{\rho}^{k}\,G^{(k)}_1(\tau,m,\epsilon_\pm)\, , \label{SymProd}
\end{align}
where $G^{(k)}_1$ can be expressed in terms of the contribution of a single M2-brane winding once around $\mathbb{S}^1$
\begin{align}
G^{(k)}_1=\frac{1}{k}\sum_{a=0}^{k-1}G^{(1)}_1(\tfrac{\tau+a}{k},m,\epsilon_\pm)\,. \label{NM1Reduction}
\end{align}
Here, $G_1^{(1)}$ can be written as the following quotient of Jacobi theta-functions
\begin{align}
G^{(1)}_1(\tau,m,\epsilon_\pm)=\frac{\theta_1(\tau;m+\epsilon_-)\theta_1(\tau;m-\epsilon_-)}{\theta_1(\tau;\epsilon_++\epsilon_-)\theta_1(\tau;\epsilon_+-\epsilon_-)}\,.
\end{align}
We see from the Fourier expansion of $G^{(k)}_1$
\bea
G^{(k)}_1=\sum_{n,\ell,r,s}c_{k}(n,\ell,r,s)\,e^{2\pi i \tau\,n}\,Q_{m}^{\ell}\,q^{r}\,t^{s}\,
\eea
(where $q=e^{i\epsilon_1}$ and $t=e^{-i\epsilon_2}$) that relation (\ref{NM1Reduction}) implies
\bea
c_{k}(n,\ell,r,s)=c_{1}(k\,n,\ell,r,s)\,.
\eea
In fact, the relation (\ref{NM1Reduction}) reflects that ${\cal Z}_{1,1}$ is the partition function of a two-dimensional ${\cal N} = (2,2)$ supersymmetric sigma model whose target space is a symmetric product, \emph{i.e.} it is the partition function of a symmetric orbifold theory and therefore it can be expressed as \cite{Dijkgraaf:1996xw}
\bea\nonumber
{\cal Z}_{1,1}&=& \prod_{k,n,\ell,r,s}(1-Q_{\rho}^{k}Q_{\tau}^{n}Q_{m}^{\ell}q^{r}t^{s})^{-c_{1}(k\,n,\ell,r,s)}\nonumber\\
&=& \sum_{k=0}^{\infty}Q_{\rho}^{k}\,\mathfrak{Z}((\mathbb{C}^2)^k/S_k), \label{FourierZ1}
\eea
where $\mathfrak{Z}((\mathbb{C}^2)^k/S_k)$ is the (equivariantly regularised) elliptic genus of $(\mathbb{C}^2)^{k}/S_{k}$ which can be written as
\bea
\mathfrak{Z}((\mathbb{C}^2)^k/S_k)=\frac{1}{k!}\sum_{{hg=gh}\atop{g,h\in S_{k}}}\sq{g}{h}\,.\label{SigmaModRep}
\eea
In the sum, \,  $\sq{g}{h}$\, denotes the ${\cal N} = (2,2)$ partition function of a sigma model whose target space is the product of $k$ copies of $\mathbb{C}^{2}$ and whose worldsheet boundary conditions in the space and time directions are twisted by $h$ and $g$, respectively. It follows from eqs.(\ref{SymProd}) and (\ref{NM1Reduction}) that
\bea
\mbox{ln}({\cal Z}_{1,1})=\sum_{k\geq 1}Q_{\rho}^{k}{\cal H}_{k}\big(G^{(1)}_1\big)\,. \label{HeckeDefZ1}
\eea
Here, ${\cal H}_{k}$ denotes the $k$-th Hecke operator: if $f_{w,\vec{r}}(\tau,\vec{z})$ is a Jacobi form of weight $w$ and index $\vec{r}$ with respect to the (multi-)argument $\vec{z}$, then the $k$-th Hecke transformation of $f_{w,\vec{r}}$ is defined as ($k\in \mathbb{N}$)
\begin{align}
&\mathcal{H}_k(f_{w,\vec{r}}(\tau,\vec{z}))=k^{w-1}\sum_{{d|k}\atop{b\text{ mod }d}}d^{-w}\,f_{w,\vec{r}}\left(\frac{k\tau+bd}{d^2},\frac{k\vec{z}}{d}\right)\,.\nonumber
\end{align}
The function $\mathcal{H}_k(f_{w,\vec{r}}(\tau,\vec{z}))$ in turn is a Jacobi form of index $r\vec{k}$ and weight $w$. Notice that,  under $SL(2,\mathbb{Z})$ transformations, $G^{(1)}_1$ transforms as
\begin{align}
G^{(1)}_1\left(-\tfrac{1}{\tau},\tfrac{m}{\tau},\tfrac{\epsilon_\pm}{\tau}\right)=e^{\frac{2\pi i}{\tau}(m^2-\epsilon_+^2)}\,G_1^{(1)}(\tau,m,\epsilon_\pm)\,,\nonumber
\end{align}
\emph{i.e.} it has weight $w=0$ and index $\vec{k}=(1,-1,0)$ with respect to $(m,\epsilon_+,\epsilon_-)$.

With eqs. (\ref{FourierZ1}), (\ref{SigmaModRep}) and (\ref{HeckeDefZ1}), we have three representations of ${\cal Z}_{1,1}$ all of which follow from each other and are a consequence of the relation (\ref{NM1Reduction}). In the following, we shall find a generalisation of the latter for the particular BPS subsector contributing to $\mathcal{F}_{N,M}$ (\ref{CalFDef}) in more complicated M-brane configurations:
%\bea
%{\cal Z}_{1,1}&=&\prod_{k,n,\ell,r,s}(1-Q_{\rho}^{k}Q_{\tau}^{n}Q_{m}^{\ell}q^{r}t^{s})^{-c_{1}(k\,n,\ell,r,s)}\\\nonumber
%&=&\sum_{k\geq 0}Q_{\rho}^{k}\,\Big(\frac{1}{k!}\sum_{gh=hg,(g,h)\in S_{k}\times S_{k}}\sq{g}{h}\Big)\\\nonumber
%&=&\mbox{exp}\Big(\sum_{k\geq 1}Q_{\rho}^{k}{\cal H}_{k}(G^{(1)})\Big)
%\eea
indeed, for arbitrary $(N,M)$, a specific part of the partition function can be identified with the elliptic genus of a product of instanton moduli spaces and can again be written as the partition function of a symmetric orbifold theory. The free energy ${\cal F}_{N,M}$ (\ref{CalFDef}) is given in terms of $G_M^{(k,\cdots,k)}$ by
\bea
{\cal F}_{N,M}&=&\sum_{n\geq 1}\frac{1}{n}\sum_{k\geq 1}Q_{\rho}^{nk}G^{(k,\cdots,k)}_M(n\tau,nm,n\epsilon_\pm)\\\nonumber
%&=&\sum_{n,k\geq 1}\frac{1}{n}Q_{\rho}^{nk}G^{(k,\cdots,k)}(n\tau,nm,n\epsilon_{1,2})\\\nonumber
&=&\sum_{K\geq 1}Q_{\rho}^{K}\sum_{n|K}\frac{1}{n}G_M^{(K/n,\cdots,K/n)}(n\tau,nm,n\epsilon_\pm)\,.
\eea
The functions $G_M^{(k,\cdots,k)}(\tau,m,\epsilon_\pm)$ diverge in the limit $\epsilon_{1,2}\mapsto 0$ proportional to $\frac{1}{\epsilon_{1}\epsilon_2}$, and so we can define the slightly modified Nekrasov-Shatashvili (NS) limit~\cite{Nekrasov:2009rc}
\bea
G_{M,\text{NS}}^{(k,\cdots,k)}=\lim_{\epsilon_2\mapsto 0}\,\tfrac{\epsilon_{2}}{\epsilon_{1}}\,G^{(k,\cdots,k)}_M(\tau,m,\epsilon_\pm)\, , \label{DefNSlimit}
\eea
which is of weight zero under modular transformations. This is in particular also the case for  $G_{M,\text{NS}}^{(1,\cdots,1)}$ which is crucial for constructing the NS-limit $\mathcal{F}_{N,M}^{\text{NS}}$ of the free energy. The functions $G_{M,\text{NS}}^{(1,\cdots,1)}$, which have weight zero and index $MN$, capture the degeneracies of monopole strings realised as M2-branes suspended between M5-branes. There is evidence that they are the (equivariantly regularised) ${\cal N} = (2,2)$ supersymmetric elliptic genera of the corresponding monopole moduli space $Y_{M,N}$ of dimension $4 MN$. For $M=1$, it was shown in \cite{Hohenegger:2015btj} that
\bea
G_{1,\text{NS}}^{(1,\cdots,1)}=N\,G^{(1)}_{1,\text{NS}}\,\Big(\frac{\theta_{+}\theta^{'}_{-}-
\theta_{-}\,\theta^{'}_{+}}{\theta_{1}(\tau;\epsilon_{1})\eta^{3}(\tau)}\Big)^{N-1}\,.
\eea
where $\theta_{\pm}=\theta_{1}(\tau;m\pm \frac{\epsilon_{1}}{2})$. The above expression is reduced to a constant $N$ for $m=\pm \frac{\epsilon_{1}}{2}$, as expected of the ${\cal N} = (2,2)$ supersymmetric elliptic genus. 
%Though we discussed only for $M=1$, 
A similar behaviour is expected for $M>1$, leading us to conjecture that $G_{M,NS}^{(1,\ldots,1)}$ is the elliptic genus of a sigma model with $\mathcal{N}=(2,2)$ supersymmetry. Hereafter, we provide evidence that, in the limit (\ref{DefNSlimit}), the degeneracies of bound states of $k$ M2-branes organise themselves as
\begin{align}
G_{M,\text{NS}}^{(k,\cdots,k)}(\tau,m,\epsilon_1)=\tfrac{1}{k}\sum_{i=0}^{k-1}G_{M,\text{NS}}^{(1,\cdots,1)}(\tfrac{\tau+i}{k},m,\epsilon_{1})\,.\label{MainResult}
\end{align}
M2-brane configurations in the NS limit are identified as monopole strings in five dimensions, so the relation \eqref{MainResult}, which constitutes the main result of this paper, is the statement that the degeneracies of bound states of monopole strings of charge $(k,\ldots,k)$ are completely determined by the degeneracies of charge $(1,\ldots,1)$ monopole strings \cite{Dijkgraaf:1996xw}. The relation (\ref{MainResult}) can equally be written in terms of Hecke transformations (also clarifying the modular transformation properties of $G_{M,\text{NS}}^{(k,\cdots,k)}$)
\begin{align}
G_{M,\text{NS}}^{(k,\cdots,k)}=\mathcal{T}_{k}(G_{M,\text{NS}}^{(1,\cdots,1)})\,,\label{Ttrafo}
\end{align}
where the operator $\mathcal{T}_{k}$ acts in the following manner on a Jacobi form $f_{w,\vec{r}}(\tau,\vec{z})$ of weight $w$ and index $\vec{r}$
\begin{align}
\mathcal{T}_{k}(f_{w,r}(\tau,\vec{z})):=\sum_{a|k}a^{w-1}\,\mu(a)\,\mathcal{H}_{\frac{k}{a}}(f_{w,\vec{r}}(a\tau,a\vec{z}))\,,\nonumber
\end{align}
and $\mu(a)$ is the M\"obius function. We can also characterise eq.(\ref{Ttrafo}) in a different fashion: If we take the Fourier expansion of $G_{M,\text{NS}}^{(1,\cdots,1)}$ as
\begin{align}\nonumber
G_{M,\text{NS}}^{(1,\cdots,1)}(\tau,m,\epsilon_1)=\sum_{p=0}^\infty\sum_{n\in\mathbb{Z}}\,c_{N,M}(p,n,r)\,Q_\tau^p\,Q_m^n\,q^{r},
\end{align}
we have the following expansion for $G_{M,\text{NS}}^{(k,\cdots,k)}$
\begin{align}
G_{M,\text{NS}}^{(k,\cdots,k)} \! =\mathcal{T}_{k}(G_{M,\text{NS}}^{(1,\cdots,1)})= \! \sum_{p=0}^\infty\sum_{n\in\mathbb{Z}}\! c_{N,M}(kp,n,r)\,Q_\tau^p\,Q_m^n\,q^r.\nonumber
\end{align}
Using the relation (\ref{MainResult}), we can express the NS-limit of the reduced free energy $\mathcal{F}_{N,M}$ (\ref{CalFDef}) as
\begin{align}
{\cal F}^{\text{NS}}_{N,M}&=\lim_{\epsilon_{2}\mapsto 0}\tfrac{\epsilon_{2}}{\epsilon_{1}}\,{\cal F}_{N,M}(\rho,\tau,m,\epsilon_1,\epsilon_2)\nonumber\\
&=\sum_{K\geq 1}Q_{\rho}^{K}\sum_{n|K}\frac{1}{n}G_{M,\text{NS}}^{(K/n,\cdots,K/n)}(n\tau,nm,n\epsilon_{1})\,,\nonumber
\end{align}
which with the relation (\ref{MainResult}) allows us to write ${\cal F}^{\text{NS}}_{N,M}$ as a sum over Hecke transformations of $G_{M,\text{NS}}^{(1,\cdots,1)}$:
\begin{align}
{\cal F}^{\text{NS}}_{N,M}&=\sum_{K\geq 1}Q_{\rho}^{K}\sum_{n|K}\frac{1}{K}\sum_{i=0}^{K/n-1}G_{M,\text{NS}}^{(1,\cdots,1)}(\tfrac{n\tau+i}{K/n},nm,n\epsilon_{1})\nonumber\\
%&=&\sum_{K\geq 1}Q_{\rho}^{K}\frac{1}{K}\sum_{n|K}\sum_{i=0}^{K/n-1}G_{NS}^{(1,\cdots,1)}(\frac{n\tau+i}{K/n},nm,n\epsilon_{1})\\
%&=&\sum_{K\geq 1}Q_{\rho}^{K}\frac{1}{K}\sum_{k|K}\sum_{i=0}^{k-1}G_{NS}^{(1,\cdots,1)}(\frac{K\,\tau+ik}{k^2},Km/k,K\epsilon_{1}/k)\\
&=\sum_{K\geq 1}Q_{\rho}^{K}{\cal H}_{K}(G_{M,\text{NS}}^{(1,\cdots,1)})\,.\label{Frewrite}
\end{align}
Using $\mathcal{F}^{\text{NS}}_{N,M}$, we can in turn define the partition function
\bea\nonumber
\tilde{\mathcal{Z}}^{NS}_{N,M}(\rho,\tau,m,\epsilon_1) \! =e^{{\cal F}_{NS}(\rho,\tau,m,\epsilon_1)}=\! \sum_{n\geq 0}\! Q_{\rho}^{n}\,\mathfrak{Z}^{N,M}_{n}(\tau,m,\epsilon_1),
\eea
which only receives contributions from the BPS sector counted by $\mathcal{F}^{\text{NS}}_{N,M}$. Here, we have
\begin{align}
\mathfrak{Z}^{N,M}_{n}(\tau,m,\epsilon_1)=\frac{1}{n!}\sum_{|\lambda|=n}a_{\lambda}\,{\cal H}_{\lambda}(G^{(1,\cdots,1)}_{M,\text{NS}}),\label{ConjClassSum}
\end{align}
where the sum is over all integer partitions $\lambda$ of $n$. Explicitly, for a partition $\lambda =(1^{m_{1}}\,2^{m_2}\,3^{m_{3}}\cdots)$, we have
\begin{align}
&a_{\lambda}=\frac{n!}{\prod_{i}m_{i}!}\,,&&{\cal H}_{\lambda}=\prod_{i}{\cal H}_{i}^{m_{i}}\,.
\end{align}
Note that the conjugacy classes of $S_n$ are equally labelled by integer partitions of $n$ such that we can rewrite the relation (\ref{ConjClassSum}) as a sum over conjugacy classes of $S_n$. Furthermore, since $\mathfrak{Z}^{N,M}_{1}(\tau,m,\epsilon_1)$ is just $G_{M,\text{NS}}^{(1,\cdots,1)}$, the elliptic genus of the target space $Y_{N,M}$, the quantity $\mathfrak{Z}_n$ is the elliptic genus of the orbifold target space $(Y_{N,M})^n/S_n$. The latter can be expressed in terms of sums over commuting pairs of elements of $S_n$ which define the twisted boundary conditions on the torus:
\bea
\mathfrak{Z}^{N,M}_{n}&=&\frac{1}{n!}\sum_{{gh=hg}\atop{g,h\in S_{n}}}\sq{g}{h}\,.
\eea
Here,  $\sq{e}{e}=\mathfrak{Z}^{N,M}_{1}$, while $\sq{g}{h}$ can be written as \cite{Bantay:1997ek}:
\begin{align}
&\sq{g}{h}=\prod_{\xi\in \mathcal{O}(g,h)}\mathfrak{Z}^{N,M}_{1}(\tau_\xi,r_\xi m,r_\xi \epsilon_1)&&\text{with} &&\begin{array}{l}g,h\in S_n\,, \\ gh=hg\,.\end{array}\nonumber
\end{align}
In this expression, $\mathfrak{s}(g,h)\subset S_n$ is the subgroup of $S_n$ generated by the (commuting pair of) elements $g,h$ and $\mathcal{O}(g,h)$ refers to the collection of orbits of $\mathfrak{s}(g,h)$ when acting on the set $\{1,\ldots,n\}$. Then, $\tau_\xi=\frac{\mu_\xi\tau+\kappa_\xi}{\lambda_\xi}$ (and $r_\xi$ is uniquely determined from the Hecke structure (\ref{ConjClassSum})), where $\lambda_\xi$ is the length of any $g$-orbit in $\xi$, $\mu_\xi$ is the number of $g$-orbits in $\xi$ (such that $\lambda_\xi\mu_\xi=|\xi|$) and $\kappa_\xi$ is the minimal non-negative integer such that $h^{\mu_\xi}=x^{\kappa_\xi}$.

It remains to provide evidence for relation (\ref{MainResult}) which is at the heart of reduced free energy (\ref{Frewrite}). We have checked the former equation by considering explicit series expansions in the parameters $Q_\tau$ and $\epsilon_1$ and comparing both sides as a function of $Q_m$. The order of $Q_\tau$ and $\epsilon$ to which we checked these relations is tabulated in Table 1.
\begin{table}[tb]
\begin{tabular}{c|c|c}
{\bf expression} & {\bf order in} $Q_\tau$ & {\bf order in} $\epsilon_1$\\\hline
&&\\[-10pt]
$G_{1,\text{NS}}^{(2)}=\mathcal{T}_2(G_{1,\text{NS}}^{(1)})$ & $20$ & $5$\\[4pt]\hline
&&\\[-10pt]
$G_{1,\text{NS}}^{(3)}=\mathcal{T}_3(G_{1,\text{NS}}^{(1)})$ & $10$ & $3$\\[4pt]\hline
&&\\[-10pt]
$G_{1,\text{NS}}^{(4)}=\mathcal{T}_4(G_{1,\text{NS}}^{(1)})$ & $5$ & $3$\\[4pt]\hline
&&\\[-10pt]
$G_{1,\text{NS}}^{(2,2)}=\mathcal{T}_2(G_{1,\text{NS}}^{(1,1)})$ & $4$ & all orders\\[4pt]\hline
&&\\[-10pt]
$G_{1,\text{NS}}^{(3,3)}=\mathcal{T}_3(G_{1,\text{NS}}^{(1,1)})$ & $2$ & all orders\\[4pt]\hline
%{\bf expression} & {\bf order in} $Q_\tau$ & {\bf order in} $\epsilon_1$\\\hline
&&\\[-10pt]
$G_{1,\text{NS}}^{(2,2,2)}=\mathcal{T}_2(G_{1,\text{NS}}^{(1,1,1)})$ & $1$ & all orders\\[4pt]\hline
&&\\[-10pt]
$G_{2,\text{NS}}^{(2,2)}=\mathcal{T}_2(G_{2,\text{NS}}^{(1)})$ & $3$ & all orders\\[4pt]\hline
&&\\[-10pt]
$G_{3,\text{NS}}^{(2,2)}=\mathcal{T}_2(G_{3,\text{NS}}^{(1)})$ & $3$  & all orders\\[4pt]
\end{tabular}
\caption{Orders to which eq.(\ref{NM1Reduction}) has been checked through explicit series expansion.}
\end{table}
This indeed provides strong evidence for relation (\ref{MainResult}); we conjecture that it holds in full generality.

To conclude, we find that the representation (\ref{ConjClassSum}) shows that the reduced partition function~$\tilde{\mathcal{Z}}_{N,M}^{\text{NS}}$ is expressible as the partition function of a symmetric orbifold CFT. Using the duality proposed in \cite{Hohenegger:2015btj}, we can further interpret $\tilde{\mathcal{Z}}_{N,M}^{\text{NS}}$ as (part of) the BPS partition function of monopole strings compactified on $\mathbb{S}^1$. Thus, our results indicate that a BPS subsector of
five-dimensional monopole strings can be written as a symmetric orbifold CFT. This CFT is a sigma model whose target space is the symmetric product of $Y_{N,M}$, where $Y_{N,M}$ is the moduli space of monopole strings with charge $(1,\ldots,1)$. Notice in this respect that, both for the interpretation of $\tilde{\mathcal{Z}}_{N,M}^{\text{NS}}$ in terms of the monopole strings (see \cite{Hohenegger:2015btj}) as well as its rewriting in terms of a symmetric orbifold partition function, the NS limit is crucial. We expect a similar orbifold description of monopole strings in little strings of $D$ and $E$ type as well.

One interesting application of our result is the counting of black hole microstates through the Ooguri-Strominger-Vafa (OSV) conjecture \cite{Ooguri:2004zv}. The little string partition function we studied is also the partition function of a subsector of the topological string on $X_{M, N}$. This implies that the black holes engineered from $X_{M, N}$ by the OSV conjecture have partition functions which in some cases can be expressed as the partition function of a sigma model. Our result indicates that perhaps in some limit this sigma model is reduced to a symmetric orbifold CFT.
%%%%%%%%%%%%%%%%%%%%%%%%%%%

%%%%%%%%%%%%%%%%%%%%%%%%%%%%%%%%%%%%%

\end{document}